\documentclass[letterpaper]{article} 
\usepackage[submission]{aaai2026}  
\usepackage{times}  
\usepackage{helvet}  
\usepackage{courier}  
\usepackage[hyphens]{url}  
\usepackage{graphicx} 
\usepackage{natbib}  
\usepackage{caption} 
\usepackage{algorithm}
\usepackage{algorithmic}
\usepackage{booktabs}
\usepackage{graphicx}
\usepackage{subcaption}
\usepackage{multirow}
\usepackage{amsmath}
\usepackage{xcolor}
\usepackage{newfloat}
\usepackage{listings}
\DeclareCaptionStyle{ruled}{labelfont=normalfont,labelsep=colon,strut=off} 
\floatstyle{ruled}
\newfloat{listing}{tb}{lst}{}
\floatname{listing}{Listing}
\title{Purely Semantic Indexing for LLM-based \\ Generative Recommendation and Retrieval}
\author{
    Ruohan Zhang,
    Jiacheng Li,
    Julian McAuley,
    Yupeng Hou
}
\affiliations{
    University of California, San Diego\\

    La Jolla, CA, United States\\
}

\usepackage{bibentry}

\begin{document}

\maketitle

\begin{abstract}
Semantic identifiers (IDs) have proven effective in adapting large language models for generative recommendation and retrieval. However, existing methods often suffer from semantic ID conflicts, where semantically similar documents (or items) are assigned identical IDs. A common strategy to avoid conflicts is to append a non-semantic token to distinguish them, which introduces randomness and expands the search space, therefore hurting performance. In this paper, we propose \textbf{purely semantic indexing} to generate unique, semantic-preserving IDs without appending non-semantic tokens. We enable unique ID assignment by relaxing the strict nearest-centroid selection and introduce two model-agnostic algorithms: exhaustive candidate matching (ECM) and recursive residual searching (RRS). Extensive experiments on sequential recommendation, product search, and document retrieval tasks demonstrate that our methods improve both overall and cold-start performance, highlighting the effectiveness of ensuring ID uniqueness. Code is available at \url{https://github.com/wangshanyw/PurelySemanticIndexing}.
\end{abstract}

\maketitle

\section{Introduction}
Semantic IDs refer to a few discrete tokens that jointly index a document (or item) in generative retrieval (or recommendation) systems~\cite{tay2022transformer,wang2022neural,rajput2023recommender}. A key property is that semantically similar documents often share the same initial digits. Training autoregressive models on these IDs allows semantically similar items to have similar generation probabilities, enabling the effective adaptation of large language models (LLMs) for generative retrieval~\cite{li2024summarization, kuo2024survey, zheng2024adapting}.

\begin{figure}[!t]
  \centering
  \includegraphics[width=\linewidth]{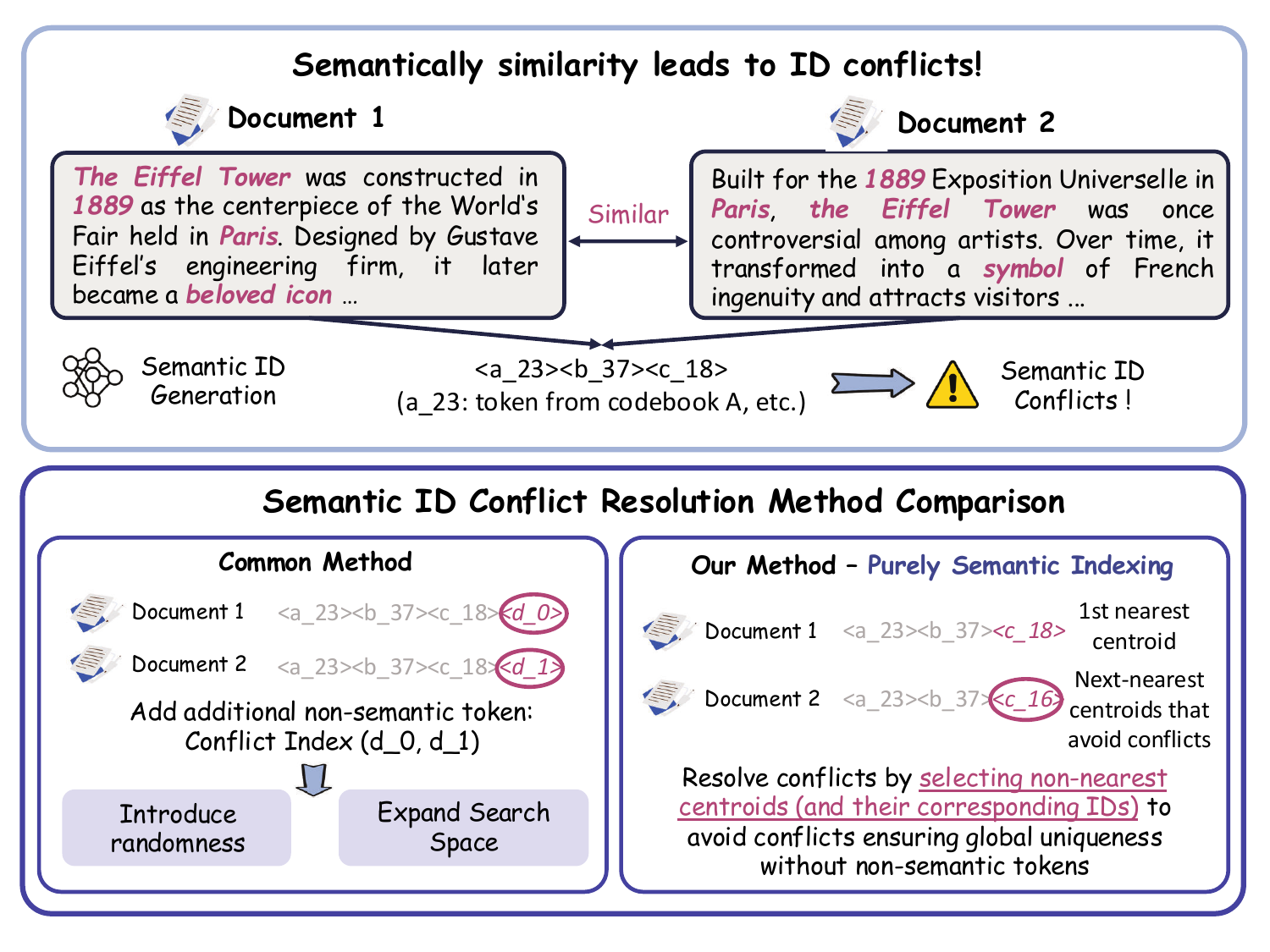}
  \caption{Illustration of the motivation and key idea of our proposed purely semantic indexing. (a) When using semantic indexing for ID construction, multiple semantically similar documents (or items) may be assigned identical IDs, leading to ID conflicts. (b) A common solution to resolve conflict is to append an additional non-semantic token (\emph{conflict index}) as a fallback. However, this introduces non-semantic randomness and expands the search space, degrading the performance. (c) To build a semantic identification system without using non-semantic tokens, we propose purely semantic indexing, which directly generates unique, semantic-preserving IDs. Our approach largely follows standard methods but allows the selection of non-nearest centroids (and their corresponding IDs) to resolve conflicts without appending additional tokens.}
  \label{fig:idea}
\end{figure}

However, when using this method for ID construction, semantic ID conflicts occur, where multiple semantically similar documents are assigned identical IDs. Although contrastive objectives~\cite{si2023generative,wang2024learnable} or regularization~\cite{wang2011regularized} during training reduce such conflicts, some identical IDs typically remain. To handle this, a common strategy is to append an additional non-semantic token (\emph{conflict index}) as a fallback to distinguish between semantically similar documents. For example, one approach is to set this token based on the number of occurrences of the same semantic prefix~\cite{rajput2023recommender}.

While this fallback approach is effective at avoiding conflicts, it introduces several drawbacks, as shown in Figure~\ref{fig:idea}. The additional tokens lack semantic meaning, introducing non-semantic randomness into an identification system originally designed to be semantic. Moreover, the increased number of tokens per document expands the search space, which complicates the retrieval process, particularly in cold-start settings where unseen tokens exacerbate generalization challenges.

\begin{figure}[t]
  \centering
  \includegraphics[width=\linewidth]{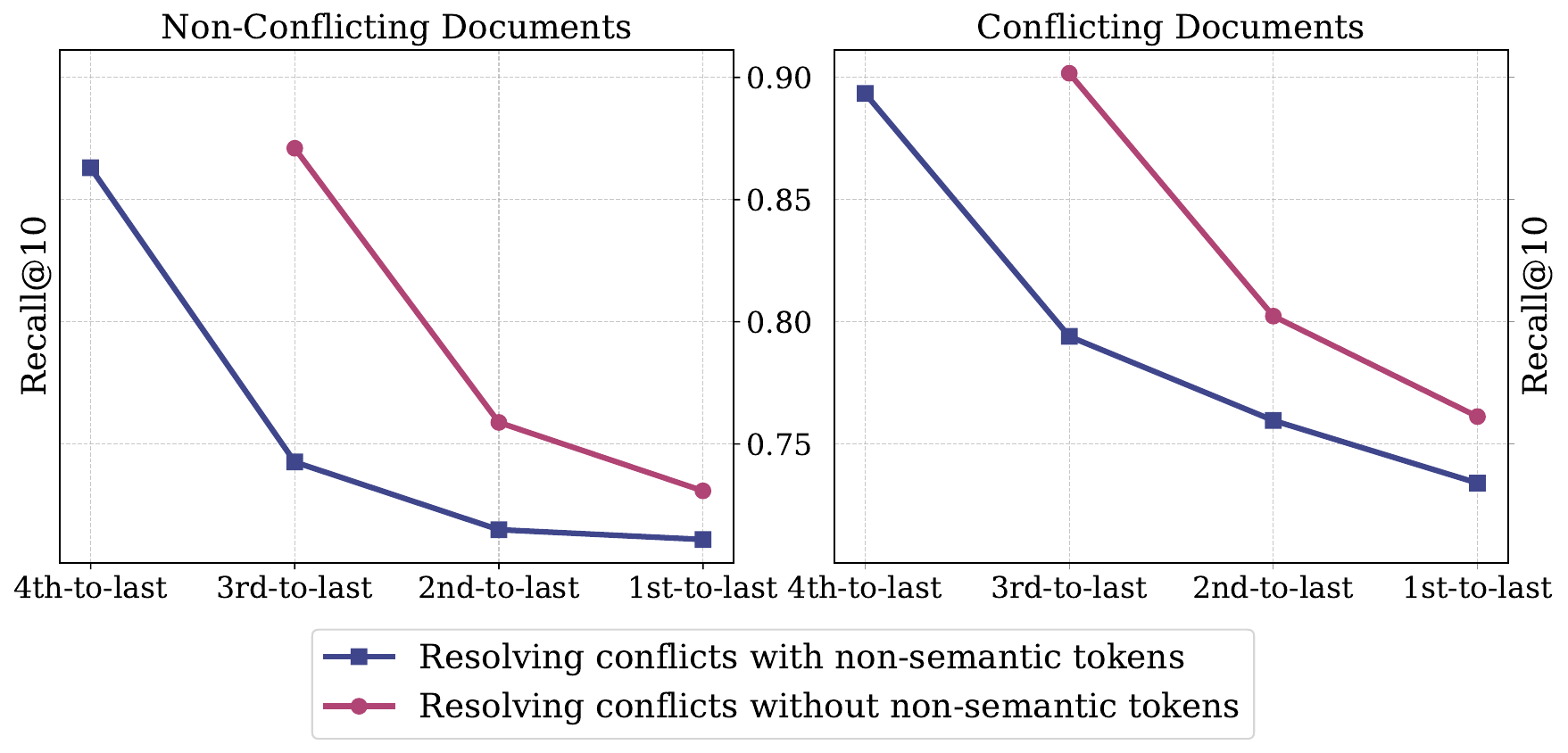}
  \caption{Retrieval performance across semantic ID levels for non-conflicting and conflicting documents on NQ320k.}
  \label{fig:motivation-analysis}
\end{figure}

As shown in our motivation study~(Figure~\ref{fig:motivation-analysis}), predicting the final non-semantic token leads to a performance drop, even when the first three semantic tokens are predicted correctly, due to the larger search space. The performance gap is especially large for conflicting documents, where relying on non-semantic tokens significantly reduces Recall@10. Additionally, the figure shows that even for documents with non-conflicting semantic prefixes, appending a non-semantic token still causes a slight performance drop, likely due to the introduced randomness. These findings validate the aforementioned drawbacks of resolving conflicts through non-semantic suffixes.

Motivated by this, we aim to develop a semantic identification system that avoids using non-semantic tokens, illustrated in Figure~\ref{fig:idea}. Our key insight is that preserving the exact correspondence or reconstructability between semantic IDs and their original semantic features is less critical than ensuring the uniqueness of the IDs. During the construction process of legacy semantic IDs, each level corresponds to a cluster center (centroid) of the input feature vectors. Rather than always assigning the ID based on the nearest centroid, we allow assignments to the second-nearest or even more relaxed choices to avoid conflicts.

To this end, we introduce \textbf{purely semantic indexing}, which directly generates unique, semantic-preserving IDs without appending additional tokens to resolve conflicts. The semantic ID construction process mostly follows typical existing methods but allows the selection of non-nearest centroids (and their corresponding IDs) to avoid conflicts. Based on different selection strategies, we introduce two variants: (1) exhaustive candidate matching (ECM), which treats each candidate semantic ID as a complete sequence, computes its distance as a whole, and selects the optimal possible one; (2) recursive residual search (RRS), which determines each token of the semantic ID recursively, layer by layer. Note that both algorithms are model-agnostic and operate with any multi-level quantization~\cite{lee2022autoregressive,van2017neural,huang2024pq} or clustering methods~\cite{murtagh2012algorithms} that produce centroid-based codebooks.

We summarize our contributions as follows:
\begin{itemize}
    \item We highlight the limitations of existing approaches that rely on non-semantic fallback tokens for conflict resolution, showing their negative impact on preserving semantics and overall performance.
    \item We propose purely semantic indexing, which directly generates unique, conflict-free semantic IDs without appending non-semantic tokens. Our key insight is that ensuring ID uniqueness is more critical than preserving an exact correspondence with the original semantic features.
    \item We conduct extensive experiments on three tasks including sequential recommendation, product search, and document retrieval. Experimental results show that our method achieves strong overall and cold-start performance, outperforming prior semantic ID strategies.
\end{itemize}
\section{Related Work}

\begin{table*}[t]
\centering
\caption{Notation used for purely semantic indexing.}
\label{tab:notation}
\begin{tabular}{l@{\hspace{0.7in}}l}
\toprule
\textbf{Symbol} & \textbf{Definition} \\
\midrule
$\mathcal{E}$ & Set of input embeddings representing documents or items \\
$L$ & Number of quantization levels in the trained RQ-VAE model \\
$\mathcal{C}^{(l)}$ & Codebook for quantization level $l$, where $l = 1, 2, \dots, L$ \\
$\text{Alloc}(\cdot)$ & Allocation function that retrieves nearest centroids \\
$k_l$ & Number of nearest centroid candidates retrieved at level $l$ \\
$\mathbf{k}$ & Top-$k$ vector: $\mathbf{k} = (k_1, k_2, \dots, k_L)$ \\
$\mathcal{C}_p$ & Set of conflict-free, purely semantic identifiers for all embeddings \\
\bottomrule
\end{tabular}
\end{table*}

\subsection{Semantic Indexing in Generative Recommendation and Retrieval}
Semantic indexing is widely used in generative retrieval~\cite{tay2022transformer, sun2023learning, wu2024generative, zhou2022ultron} and recommendation~\cite{rajput2023recommender,hua2023index,wei2023towards, tan2024idgenrec} to represent document or item semantic content for its ability to index large volumes of documents (or items) using a compact vocabulary~\cite{gen-survey, rajput2023recommender}. A variety of semantic indexing approaches have been proposed to generate discrete identifiers.

\subsection{Semantic ID Construction Techniques}
Semantic ID construction techniques can be broadly categorized into the following types:

\begin{itemize}
    \item \textit{Vector Quantization.} 
    Vector quantization is a widely adopted method for constructing semantic IDs~\cite{zhou2022ultron,hou2023learning, zhu2024cost, wang2024content, sun2023learning, petrov2023generative,liu2024mbgen,lin2025order,hou2025rpg}. It encodes the concatenated document (or item) features into vector representations using a pre-trained language model such as Sentence-T5~\cite{ni2021sentence}, and then quantizes them into a fixed-length sequence of discrete tokens using quantization techniques like RQ-VAE~\cite{lee2022autoregressive}, etc.

    \item \textit{Hierarchical Clustering.} Hierarchical clustering–based methods organize embeddings into tree structures to enable scalable search, clustering documents (or items) at multiple levels, as demonstrated in DSI~\cite{tay2022transformer}, NCI~\cite{wang2022neural}, and SEATER~\cite{si2023generative}. The resulting semantic IDs often have variable lengths and offer greater interpretability~\cite{hua2023index}.

    \item \textit{Language Model-based ID Generator}
    Recent approaches leverage language models to generate semantic IDs. LMIndexer~\cite{jin2023language} unifies ID generation and retrieval with end-to-end training, where one model generates semantic IDs from input text and another reconstructs the input in a self-supervised manner. IDGenRec~\cite{tan2024idgenrec} extracts natural language tokens from the input features and uses a concise textual identifier to represent each item as the semantic ID.
    
    \item \textit{Context-aware Tokenization}
    Context-aware tokenization has become a popular approach for capturing contextual information within input interaction sequences. ActionPiece~\cite{hou2025actionpiece} introduces a context-aware tokenization for user action sequences. Rather than assigning the same token to an item every time, it represents each user action as a set of item features and constructs tokens by merging feature patterns that frequently co-occur in context.
\end{itemize}

In this work, we propose purely semantic indexing, which generates unique, meaning-preserving IDs without appending non-semantic tokens. Our approach is compatible with existing quantization methods like RQ-VAE~\cite{lee2022autoregressive} and hierarchical clustering methods~\cite{murtagh2012algorithms}.

\section{Proposed Approach}
\subsection{Problem Setting}

To address semantic ID conflicts without relying on non-semantic tokens, we consider a semantic indexing scenario where the goal is to assign a unique, fully-semantic, and conflict-free identifier to each document (or item) based on its semantic content. Table~\ref{tab:notation} summarizes notations used throughout this section. The objective is to assign each embedding $e \in \mathcal{E}$ a unique and meaningful identifier that satisfies:
\begin{itemize}
    \item The assigned ID does not conflict with any other one.
    \item The ID reflects codebook-learned semantic structures.
    \item No non-semantic tokens are introduced as a fallback.
\end{itemize}

We define the final mapping as $\mathcal{C}_p: \mathcal{E} \rightarrow \mathcal{C}^{(1)} \times \cdots \times \mathcal{C}^{(L)}$, where each code is selected from the corresponding candidate lists produced by $\text{Alloc}$. The key challenge lies in resolving potential ID conflicts while preserving semantics. 

Therefore, we propose purely semantic indexing. While our ID construction process largely follows existing methods, it allows for the selection of second-nearest or more relaxed centroid assignments to resolve conflicts. In the following subsections, we introduce two specific algorithms that are compatible with any multi-level quantization~\cite{lee2022autoregressive,van2017neural,huang2024pq} or clustering method~\cite{murtagh2012algorithms} based on centroid-based codebooks. For clarity, we illustrate the approach using RQ-VAE~\cite{lee2022autoregressive} as the base indexing method. 

\subsection{Exhaustive Candidate Matching (ECM)}
ECM is designed to resolve semantic ID conflicts by globally optimizing the ID assignment process. Instead of greedily selecting the nearest centroid at each level, ECM considers multiple top-ranked candidates per layer and exhaustively evaluates all possible combinations of these candidates across layers. Each combination is treated as a complete semantic ID sequence, and a global score is computed for each. Among all valid, unassigned sequences, the one with the best score is selected.

As shown in the Alg.~\ref{alg:ECM}, we assume a set of input embeddings $\mathcal{E}$ representing documents (or items), along with a trained RQ-VAE model consisting of $L$ quantization levels and corresponding codebooks ${\mathcal{C}^{(l)}}_{l=1}^{L}$. For each level $l$, a list of the $k_l$ nearest candidate centroids is retrieved using an allocation function $\text{Alloc}(\cdot)$, where the top-$k$ vector is $\mathbf{k} = (k_1, k_2, \dots, k_L)$. The goal is to construct a conflict-free set of purely semantic identifiers $\mathcal{C}_p$ for all embeddings by selecting unique and meaningful combinations of tokens across the $L$ quantization levels.

ECM first obtains the candidate centroids and residuals at all levels. For each embedding, all combinations with $k_l$ nearest centroids selected at each level $l$ are enumerated to form candidate IDs. Each candidate ID $\mathbf{c}$ is evaluated by the negative sum of residual norms:

\begin{equation}
\text{score}(c) = -\sum_{l=1}^{L} \left\| 
\text{residuals}^{(l)}_{c^{(l)}} \right\|_2
\label{eq:score_function}
\end{equation}

\noindent The candidate IDs are sorted in descending order of score, and the first candidate without conflict $\mathbf{c}$ is assigned to the embedding, ensuring uniqueness throughout $\mathcal{E}$.

ECM exhaustively explores the candidate space to ensure each embedding is assigned a semantic-preserving and conflict-free ID. While ECM guarantees optimality by sorting candidates by scores, its complexity is exponential in $L$ and the $k_l$ size at each level, making it sometimes expensive as the number of quantization levels or candidates grows.

\begin{algorithm}[t]
\caption{Exhaustive Candidate Matching (ECM)}
\label{alg:ECM}
\begin{algorithmic}[1]
\STATE \textbf{Input:} Embeddings $\mathcal{E}$; codebooks $\{\mathcal{C}^{(l)}\}_{l=1}^{L}$; allocation function $\text{Alloc}(\cdot)$; top-$k$ vector $\mathbf{k}=(k_1, k_2, \dots, k_L)$; 
\STATE \textbf{Output:} Purely Semantic IDs mapping $\mathcal{C}_p$

\STATE \textbf{Initialization: }IDs mapping $\mathcal{C}_p \gets \{\}$ and used ID set $\mathcal{U} \gets \emptyset$
\FOR{$l = 1$ to $L$}
    \STATE $(\text{candidates}^{(l)}, \text{residuals}^{(l)})\ \gets\ \text{Alloc}(\mathcal{E},\ \mathcal{C}^{(l)},\ k_l)$
\ENDFOR
\FOR{each embedding $e \in \mathcal{E}$}
    \STATE candidate\_groups $\gets [\ ]$
    \FORALL{centroid combinations across $L$ levels, with $k_l$ centroids selected at each level $l$}
        \STATE candidate ID $\mathbf{c} \gets \text{Concat}(\text{centroid indices})$
        \STATE $\text{score}(\mathbf{c}) = -\sum_{l=1}^{L} \| \text{residuals}^{(l)}_{\mathbf{c}^{(l)}} \|$
        \STATE Append $(\mathbf{c}, \text{score}(\mathbf{c}))$ to candidate\_groups
    \ENDFOR
    \STATE Sort $\text{candidate\_groups}$ in \textbf{descending} order of score
    \FOR{each $\mathbf{c}$ in $\text{candidate\_groups}$}
        \IF{$\mathbf{c} \notin \mathcal{U}$}
            \STATE $\mathcal{C}_p[e] \gets \mathbf{c}$
            \STATE $\mathcal{U} \gets \mathcal{U} \cup \{\mathbf{c}\}$
            \STATE \textbf{break}
        \ENDIF
    \ENDFOR
\ENDFOR
\RETURN $\mathcal{C}_p$
\end{algorithmic}
\end{algorithm}

\subsection{Recursive Residual Searching (RRS)}
For scalability, we introduce RRS. Instead of globally evaluating all combinations, RRS constructs semantic IDs recursively by selecting centroid candidates at each layer based on local residuals, greedily forming conflict-free IDs.

As shown in Alg.~\ref{alg:RRS}, RRS initiates a recursive search for each embedding, constructing semantic IDs incrementally. At each level $l$, the algorithm utilizes the allocation function $\text{Alloc}$ to retrieve the $k_l$ nearest centroid candidates and their corresponding residuals. For each centroid, the current partial code ID is extended, and the search continues recursively to the next level using the updated residual vector. If a complete token combination (of length $L$) is constructed and has not been previously assigned, it is immediately accepted and assigned to this embedding. If a conflict is encountered, the search backtracks and continues with alternative candidates. 

RRS significantly reduces the search space and runtime compared to the full Cartesian product enumeration used in Alg.~\ref{alg:ECM}. While RRS assigns IDs greedily based on local decisions and does not guarantee globally optimal combinations, it is highly efficient and often yields valid, semantically meaningful IDs. In contrast to ECM's global computation, RRS dynamically updates residuals based on the currently selected centroid during the search, allowing it to directly assess the downstream impact of each selection. As a result, despite its narrower candidate space, RRS can yield higher-quality IDs than ECM in some cases.

\begin{algorithm}[t]
\caption{Recursive Residual Searching (RRS)}
\label{alg:RRS}
\begin{algorithmic}[1]
\STATE \textbf{Input/Output/Initialization:} Same as Alg.~\ref{alg:ECM}

\STATE \textbf{Function} \text{SearchNewID}($x$, $l$, current\_id):
\IF{$l > L$}
    \STATE \textbf{return} current\_id
\ENDIF
\STATE $(\text{centroids},\ \text{residuals}) \gets \text{Alloc}(x,\ \mathcal{C}^{(l)},\ k_l)$
\FOR{$r = 1$ to $k_l$}
    \STATE $\text{new\_id} \gets \text{Concat}(\text{current\_id},\text{centroids}_{r})$
    \STATE $\text{next\_residual} \gets \text{residuals}_{r}$
    \STATE $\text{valid\_id} \gets \text{SearchNewID}(\text{next\_residual},\ l+1,\ \text{new\_id})$
    \IF{$\text{valid\_id} \notin \mathcal{U}$}
        \STATE \textbf{return} $\text{valid\_id}$
    \ENDIF
\ENDFOR
\STATE \textbf{return} None \COMMENT{No valid ID found along this path}

\FOR{each embedding $e \in \mathcal{E}$}
    \STATE $\text{searched\_id} \gets \text{SearchNewID}(e, 1, \text{""})$
    \IF{$\exists\ \text{searched\_id}$ and $\text{searched\_id} \notin \mathcal{U}$}
        \STATE $\mathcal{C}_p[e] \gets \text{searched\_id}$
        \STATE $\mathcal{U} \gets \mathcal{U} \cup \{\text{searched\_id}\}$
    \ENDIF
\ENDFOR
\RETURN $\mathcal{C}_p$
\end{algorithmic}
\end{algorithm}

\section{Experiments}
\begin{table*}[t]
\centering
\caption{Statistics for Amazon datasets used in sequential recommendation and product search~\cite{jin2023language}.}
\label{tab:amazon_stats}
\begin{tabular}
{@{}l@{\hspace{0.26in}}c@{\hspace{0.26in}}c@{\hspace{0.26in}}c@{\hspace{0.26in}}c@{\hspace{0.27in}}c@{\hspace{0.27in}}c@{}}
\toprule
\textbf{Dataset} & \textbf{\#Items} & \textbf{\#Users} & \textbf{Rec History} & \textbf{Search Query} & \textbf{Search Labels} \\
 & & & (train/dev/test) & (train/dev/test) & (train/dev/test) \\
\midrule
Amazon-Beauty  & 12,101 & 22,363 & 111,815 / 22,363 / 22,363 & 1,049 / 150 / 338 & 1,907 / 268 / 582 \\
Amazon-Sports  & 18,357 & 35,598 & 177,990 / 35,598 / 35,598 & 1,299 / 186 / 443 & 2,209 / 311 / 764 \\
Amazon-Toys    & 11,924 & 19,412 & 97,060 / 19,412 / 19,412 & 1,010 / 145 / 351 & 1,653 / 250 / 594 \\
\bottomrule
\end{tabular}
\end{table*}

\begin{table*}[t]
\centering
\caption{Statistics for datasets used in document retrieval~\cite{jin2023language}.}
\label{tab:retrieval_stats}
\begin{tabular}{@{}l@{\hspace{0.76in}}c@{\hspace{0.76in}}c@{\hspace{0.76in}}c@{}}
\toprule
\textbf{Dataset} & \textbf{\#Documents} & \textbf{Queries (train/test)} & \textbf{Search Labels (train/test)} \\
\midrule
NQ320k     & 109,739     & 307,373 / 7,830    & 307,373 / 7,830 \\
MS MARCO 1M & 1,000,000   & 502,939 / 6,980    & 532,751 / 7,437\\
\bottomrule
\end{tabular}
\end{table*}

\begin{table*}[!t]
\centering
\caption{Training hyperparameters used across different tasks.}
\label{tab:hyperparams}
\begin{tabular}
{@{}l@{\hspace{0.2in}}c@{\hspace{0.2in}}c@{\hspace{0.2in}}c@{\hspace{0.2in}}c@{}}
\toprule
\textbf{Hyperparameter} & \textbf{Amazon-Rec} & \textbf{Amazon-Product Search} & \textbf{NQ320k} & \textbf{MS MARCO-1M} \\
\midrule
Training Steps               & 15,000  & 15,000  & 30,000  & 30,000 \\
Learning Rate           & \{1e-4,\ 5e-4,\ 1e-3\}    & \{6e-4,\ 8e-4,\ 1e-3\}    & \{2e-3,\ 5e-3,\ 1e-2\}    & \{2e-3,\ 5e-3,\ 1e-2\} \\
Total Batch Size        & 32      & 32      & 1024    & 16,384 \\
Gradient Accum. Steps   & 2       & 1       & 8       & 16 \\
Max Source Length       & 1024    & 1024    & 1024    & 32 \\
Max Target Length       & 32      & 128     & 128     & 32 \\
Num Beams               & 20      & 20      & 20      & 20 \\
Weight Decay            & 0.01    & –       & –       & 0.01 \\
\bottomrule
\end{tabular}
\end{table*}

\subsection{Downstream Tasks}
We evaluate the effectiveness of our methods in three downstream tasks, following the task definition, dataset processing, and experiment setup introduced in LMIndexer~\cite{jin2023language}: Sequential Recommendation, Product Search, and Document Retrieval. The dataset statistics for sequential recommendation and product search are presented in Table~\ref{tab:amazon_stats}, while those for document retrieval are shown in Table~\ref{tab:retrieval_stats}~\cite{jin2023language}. We integrate our proposed ECM and RSS with two base semantic ID indexers, RQ-VAE~\cite{lee2022autoregressive} and hierarchical clustering (HC)~\cite{murtagh2012algorithms, tay2022transformer}, and evaluate their performance in generative settings against the corresponding indexers without our methods. The models are initialized with the same T5 base checkpoint~\cite{2020t5}. The hyperparameters for training procedures for each task are listed in Table~\ref{tab:hyperparams}. All experiments are conducted on 2 NVIDIA RTX A6000 40GB GPUs. We detail the experiments below.

\begin{table*}[t]
\centering
\caption{Performance on sequential recommendation tasks. $*$ indicates a statistically significant improvement over RQ-VAE (or HC) with $p$-value $<$ 0.05 (two-tailed t-test over 3 trials).}
\label{tab:seq_rec}
\begin{tabular}
{@{}c@{\hspace{0.39in}}c@{\hspace{0.39in}}c@{\hspace{0.39in}}c@{\hspace{0.39in}}c@{\hspace{0.39in}}c@{\hspace{0.39in}}c@{}}
\toprule
\multirow{2.5}{*}{\textbf{Indexing Method}}& \multicolumn{2}{c}{\textbf{Amazon-Beauty}} & \multicolumn{2}{c}{\textbf{Amazon-Sports}} & \multicolumn{2}{c}{\textbf{Amazon-Toys}} \\ \cmidrule(lr){2-3} \cmidrule(lr){4-5} \cmidrule(lr){6-7}
 & \textbf{Recall@5}     & \textbf{NDCG@5}    & \textbf{Recall@5}     & \textbf{NDCG@5}    & \textbf{Recall@5}    & \textbf{NDCG@5}   \\ \midrule
\textbf{RQ-VAE}         & 0.0184                      & 0.0115                   & 0.0124                      & 0.0082                   & 0.0176                     & 0.0116                  \\
\textbf{RQ-VAE-ECM}     & \textbf{0.0191}*                     & 0.0116*          & 0.0132*                       & 0.0090*                   & 0.0183*                     & 0.0117*                  \\
\textbf{RQ-VAE-RRS}     & 0.0187*                      & \textbf{0.0118}*                   & \textbf{0.0133}*                      & \textbf{0.0091}*                   & \textbf{0.0188}*                     & \textbf{0.0119}*                  \\ \midrule
\textbf{HC}            & 0.0190                      & 0.0118                   & 0.0106                      & 0.0066                   & 0.0171                      & 0.0110                  \\
\textbf{HC-ECM}        & \textbf{0.0203}*                      & \textbf{0.0128}*                    & 0.0122*                      & 0.0074*                   & 0.0175*                     & 0.0109*                  \\
\textbf{HC-RRS}        & 0.0199*                      & 0.0124*                    & \textbf{0.0126}*                      & \textbf{0.0080}*                    & \textbf{0.0176}*                      & \textbf{0.0116}*                   \\ \bottomrule
\end{tabular}
\end{table*}

\begin{table*}[t]
\centering
\caption{Performance on product search tasks. * denotes significance as defined in Table ~\ref{tab:seq_rec}.}
\label{tab:product}
\begin{tabular}
{@{}c@{\hspace{0.42in}}c@{\hspace{0.42in}}c@{\hspace{0.42in}}c@{\hspace{0.42in}}c@{\hspace{0.42in}}c@{\hspace{0.42in}}c@{}}
\toprule
\multirow{2.5}{*}{\textbf{Indexing Method}} & \multicolumn{2}{c}{\textbf{Amazon-Beauty}} & \multicolumn{2}{c}{\textbf{Amazon-Sports}} & \multicolumn{2}{c}{\textbf{Amazon-Toys}} \\ \cmidrule(lr){2-3} \cmidrule(lr){4-5} \cmidrule(lr){6-7}
& \textbf{NDCG@5}      & \textbf{MAP@5}      & \textbf{NDCG@5}      & \textbf{MAP@5}      & \textbf{NDCG@5}     & \textbf{MAP@5}     \\ \midrule
\textbf{RQ-VAE}         & 0.2611                     & 0.2364                    & 0.2804                     & 0.2562                    & 0.2495                    & 0.2278                   \\
\textbf{RQ-VAE-ECM}     & 0.2692*                     & 0.2445*           & 0.2863*                     & 0.2624*                    & \textbf{0.2694}*                    & \textbf{0.2486}*                   \\
\textbf{RQ-VAE-RRS}     & \textbf{0.2702}*                     & \textbf{0.2452}*                    & \textbf{0.2870}*                     & \textbf{0.2629}*                    & 0.2624*                    & 0.2413*                   \\ \midrule
\textbf{HC}            & 0.2333                     & 0.2105                    & 0.2405                     & 0.2193                    & 0.2618                    & 0.2356                    \\
\textbf{HC-ECM}        & \textbf{0.2436}*                     & 0.2179*                    & 0.2651*                     & 0.2415*                    & \textbf{0.2671}*                    & \textbf{0.2440}*                   \\
\textbf{HC-RRS}        & 0.2435*                     & \textbf{0.2234}*                    & \textbf{0.2783}*                     & \textbf{0.2551}*                    & 0.2623*                     & 0.2371*                    \\ \bottomrule
\end{tabular}
\end{table*}

\begin{table*}[!t]
\begin{minipage}[!t]{0.55\textwidth}
\centering
\small
\caption{Performance on document retrieval tasks. * denotes significance as defined in Table ~\ref{tab:seq_rec}.}
\label{tab:doc}
\begin{tabular}{@{}ccccc@{}}
\toprule
\multirow{2.5}{*}{\textbf{Indexing Method}}& \multicolumn{2}{c}{\textbf{NQ320k}}    & \multicolumn{2}{c}{\textbf{MS MARCO-1M}} \\ \cmidrule(lr){2-3} \cmidrule(lr){4-5}
 & \textbf{Recall@1} & \textbf{Recall@10} & \textbf{Recall@1}  & \textbf{Recall@10} \\ \midrule
\textbf{RQ-VAE}         & 0.6318                  & 0.8332                   & 0.1965                    & 0.4329                   \\
\textbf{RQ-VAE-ECM}     & \textbf{0.6447}*                  & \textbf{0.8478}*          & \textbf{0.2101}*                   & \textbf{0.4415}*                   \\
\textbf{RQ-VAE-RRS}     & 0.6446*                  & 0.8360*                   & 0.2071*                   & 0.4365*                   \\ \midrule
\textbf{HC}            & 0.5484                  & 0.7458                   & 0.1821                   & 0.4235                   \\
\textbf{HC-ECM}        & \textbf{0.5503}*                  & \textbf{0.7492}*                   & \textbf{0.1896}*                   & \textbf{0.4288}*                   \\
\textbf{HC-RRS}        & 0.5496*                  & 0.7479*                   & 0.1864*                    & 0.4266*                   \\ \bottomrule

\end{tabular}
\end{minipage}%
\hfill
\begin{minipage}[!t]{0.4\textwidth}
\centering
\caption{Ablation results showing the impact of the number of ID levels on Amazon-Sports product search.}
\label{tab:id_levels}
\begin{tabular}{@{}lcc@{}}
\toprule
\textbf{ID Scheme} & \textbf{NDCG@5} & \textbf{MAP@5} \\
\midrule
Two-level + Conflict Idx & 0.2629 & 0.2402 \\
Purely Semantic (ECM) & 0.2863 & 0.2624 \\
Purely Semantic (RRS) & \textbf{0.2870} & \textbf{0.2629} \\
\bottomrule
\end{tabular}
\end{minipage}
\end{table*}

\subsection{Sequential Recommendation}
\label{sec:seq}

\noindent \textbf{Task Definition.} Given a user's historical interaction sequence $I_u$, the task is to predict the next item $v$ with which the user interacts.

\noindent \textbf{Datasets.} We use three domains from the Amazon review dataset~\cite{he2016ups}: \textit{Beauty}, \textit{Sports}, and \textit{Toys}. Users and items with at least five interactions are retained. For each user, the last interaction is used for testing, the second-to-last interaction is used for validation, and the remaining interactions are used for training.

\noindent \textbf{Implementation Details.} 
We evaluate our methods on three domains from the Amazon review dataset~\cite{he2016ups}: \textit{Beauty}, \textit{Sports}, and \textit{Toys}. All models are trained for $15, 000$ steps, with a batch size of 32, and a maximum input length of 1024 tokens. The learning rate is searched over $\{$1e-4$,\ $5e-4$,\ $1e-3$\}$, which is listed in Table~\ref{tab:hyperparams}.

\noindent \textbf{Result.} The performance comparisons are shown in Table~\ref{tab:seq_rec}. We observe that both ECM and RRS consistently improve performance over the vanilla indexing methods across all domains. The results highlight improving the quality of semantic indexing can enhance the performance of generative sequential recommendation.

\subsection{Product Search}

\noindent \textbf{Task Definition.} Given a query $q$, retrieve an item $v$ from the corpus that matches the user's intent.

\noindent \textbf{Datasets.} We use the Amazon Product Search dataset~\cite{reddy2022shopping}, focusing on the same three domains: \textit{Beauty}, \textit{Sports}, and \textit{Toys}. We retain only queries that correspond to ground-truth products in the corpus used in Section~\ref{sec:seq} and adopt the original train/test split. 

\noindent \textbf{Implementation Details.} We use the Amazon Product Search dataset~\cite{reddy2022shopping}, focusing on the same domains in Section~\ref{sec:seq}. All models are trained for $15,000$ steps, with a learning rate grid of $\{$6e-4$,\ $8e-4$,\ $1e-3$\}$. The batch size is 32, and the input length limit is 1024.

\noindent \textbf{Result.} The performances are shown in Table~\ref{tab:product}. ECM and RRS boost the performance of vanilla indexing methods in all three domains. Compared with the Beauty and Toys domains, the Sports domain shows greater improvement, which may be attributed to higher semantic content density of items. Resolving ID conflicts by purely semantic indexing improves small-scale generative retrieval performance.

\subsection{Document Retrieval}

\noindent \textbf{Task Definition.} Given a query $q$, retrieve relevant documents $v$ from a large document collection.

\noindent \textbf{Datasets.} We evaluate on \textit{Natural Questions (NQ)}~\cite{kwiatkowski2019natural} and \textit{MS MARCO}~\cite{nguyen2016ms}. We used the constructed subset named \textit{MS MARCO-1M}~\cite{pradeep2023does, jin2023language} here. 

\noindent \textbf{Implementation Details.} We evaluate on \textit{Natural Questions (NQ 320k)}~\cite{kwiatkowski2019natural} and \textit{MS MARCO-1M}~\cite{nguyen2016ms, pradeep2023does, jin2023language}. The training sets are augmented using \textit{docT5query}~\cite{nogueira2019doc2query, wang2022neural}. Models are trained for $30,000$ steps, with learning rates searched over $\{$2e-3$,\ $5e-3$,\ $1e-2$\}$. The batch size is set to $1024$ for NQ and $16384$ for MS MARCO, with a maximum input text length of $1024$ for NQ and $32$ for MS MARCO.

\noindent \textbf{Result.} The retrieval performance is reported in Table~\ref{tab:doc}. We find that both ECM and RRS enhance the vanilla indexing me across both datasets, demonstrating their effectiveness in large-scale generative retrieval. Improvements on MS MARCO are smaller than NQ, likely due to its larger document pool and higher retrieval difficulty. Large-scale datasets still require improved semantic indexing methods.

\subsection{Experimental Stability and Significance}  
We conduct three independent trials per setting. We report the average performance across trials and include statistical significance tests and corresponding p-values for all pairwise comparisons. As denoted in Table~\ref{tab:seq_rec},~\ref{tab:product},~\ref{tab:doc}, both ECM and RRS significantly outperform vanilla RQ-VAE (or HC) in most settings (e.g., $p < 0.05$, often $p < 0.02$). Variance across runs remains consistently low, demonstrating the stability of our proposed methods.

\subsection{ECM vs. RRS Behavior}
On datasets with high centroid overlap and limited conflict diversity, RRS’s recursive assignment strategy can avoid redundant trials and yield better solutions (e.g., Amazon-Sports). In contrast, ECM tends to perform better when exhaustive exploration of diverse candidates is required (e.g., NQ320k), where broader semantic variation benefits from global scoring. This suggests the two methods are complementary, with RRS excelling in structured spaces and ECM in more diverse or ambiguous ones.

\subsection{Ablation Study}
To analyze the design choices and robustness of our approach, we conduct two ablation studies: (1) the number of ID levels and (2) the candidate ranking strategy in ECM.

\subsubsection{Number of ID Levels} We compare models with three-level purely semantic IDs to those using two levels plus a non-semantic conflict index on the Amazon product search task (codebook size 256). As shown in Table~\ref{tab:id_levels}, fully semantic IDs (three levels) consistently outperform IDs based on the conflict index, demonstrating the benefits of preserving the semantic structure in all ID tokens.

\subsubsection{Candidate Ranking Strategy} We evaluate three candidate ranking strategies in ECM: random selection, combination order ranking, and sorting by the negative sum of residual norms 
, which is the scoring method ultimately adopted in our approach. Table~\ref{tab:candidate_ranking} shows that our score function performs best, highlighting the importance of cumulative semantic alignment across ID levels.

\begin{table}[t]
\centering
\caption{Ablation results showing the impact of candidate ranking strategy on Amazon-Sports product search.}
\label{tab:candidate_ranking}
\begin{tabular}{@{}l@{\hspace{0.5in}}c@{\hspace{0.5in}}c@{}}
\toprule
\textbf{Ranking Strategy} & \textbf{NDCG@5} & \textbf{MAP@5} \\
\midrule
Random Selection & 0.2169 & 0.1955 \\
Combination Order & 0.2608 & 0.2389 \\
Our Score Function & \textbf{0.2694} & \textbf{0.2486} \\
\bottomrule
\end{tabular}
\end{table}

\begin{figure}[t]
  \centering
    \centering
    \includegraphics[width=\linewidth]{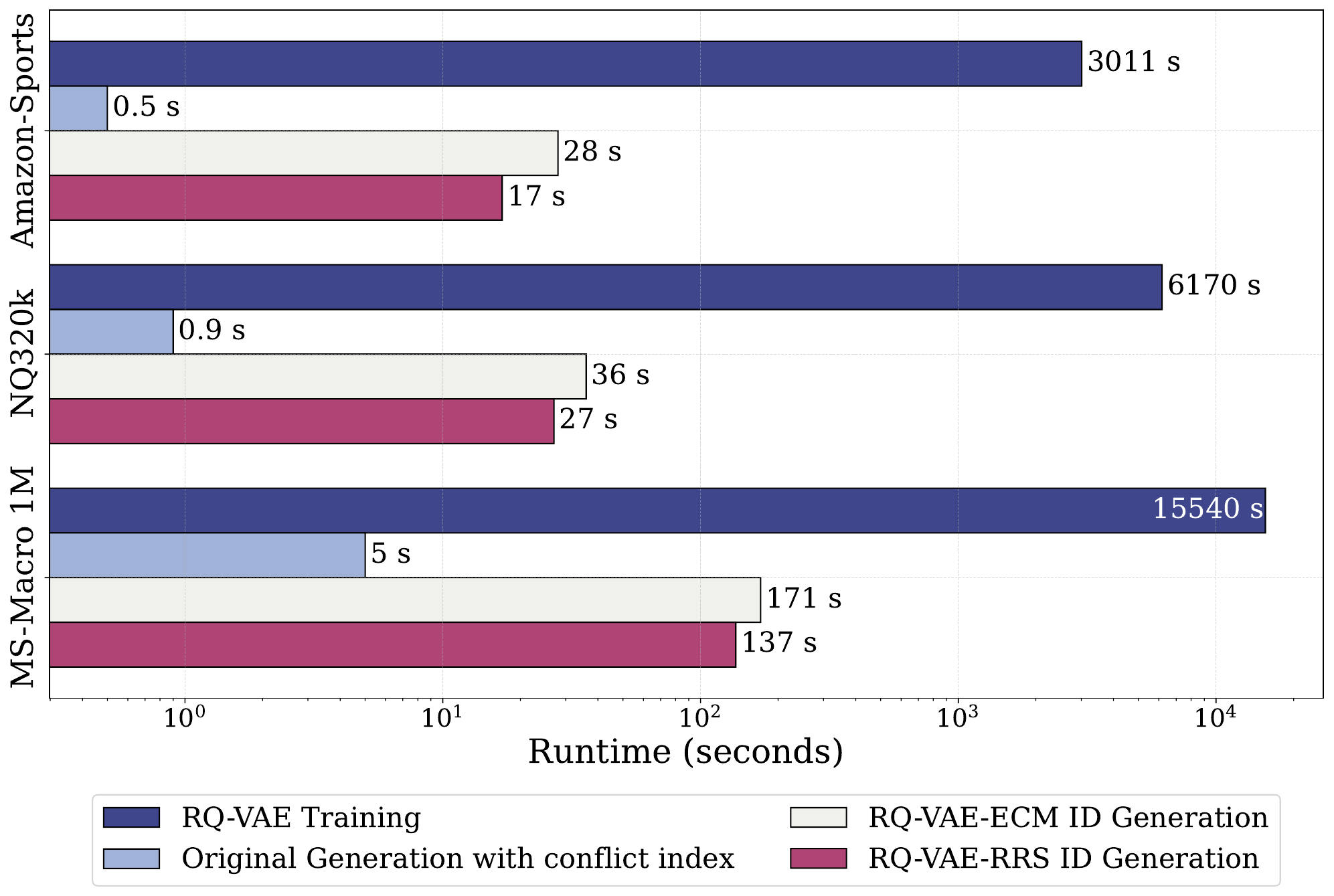}
    \caption{Runtime comparison (log scale) of RQ-VAE training and ID generation methods across datasets.}
    \label{fig:runtime-analysis}
\end{figure}

\begin{figure}[t]
  \centering
    \centering
    \includegraphics[width=0.8\linewidth]{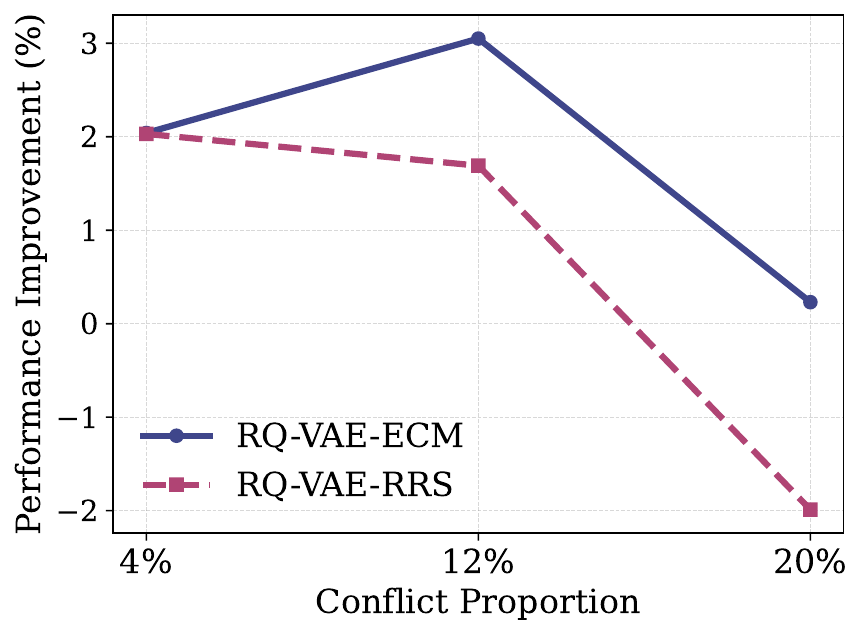}
    \caption{Impact of semantic ID conflict proportion on NQ320k retrieval performance improvement.}
    \label{fig:conflict_prop}
\end{figure}
\subsection{Runtime and Scalability Analysis}
We evaluate the runtime and scalability of our proposed method by analyzing the end-to-end cost of training and ID generation across various datasets, as summarized in Figure~\ref{fig:runtime-analysis} on a logarithmic scale. The dominant portion of the total runtime is from RQ-VAE training, which ranges from $10^4$ to $10^3$ seconds depending on dataset size, whereas the actual ID generation is comparatively lightweight, often requiring around $10^2$ seconds. Although our methods incur slightly higher ID generation costs compared with the original ID approach (e.g., 171s vs. 5s on MS MARCO; 36s and 27s vs. 0.9s on NQ320k), the overall runtime remains practical for offline indexing pipelines, particularly when using appropriately chosen model and clustering configurations. Furthermore, in real-world applications, ID generation is typically infrequent. Once an RQ-VAE model is trained, newly added items or documents can be encoded using the saved checkpoints, avoiding the need for repeated retraining and making the approach feasible for incremental updates.

\begin{figure}[t]
  \centering
    \centering
    \includegraphics[width=\linewidth]{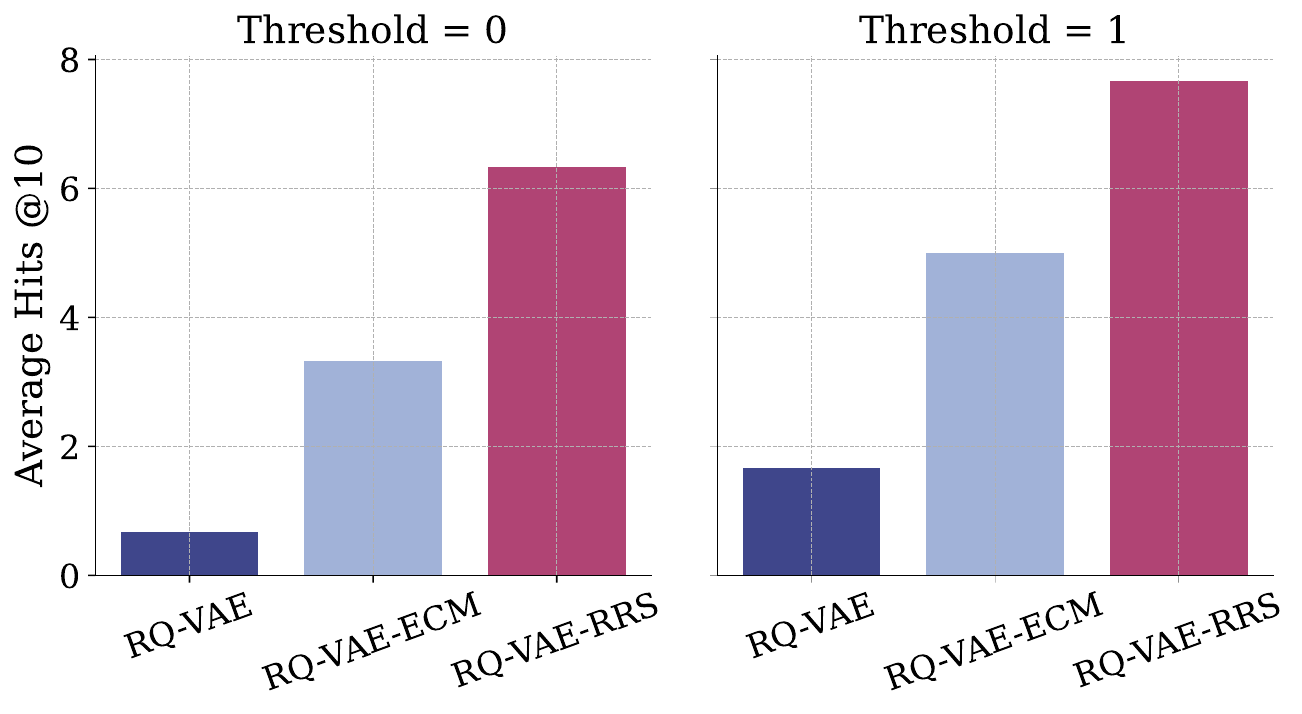}
    \caption{Average performance on cold-start items on Amazon-Toys sequential recommendation over 3 trials. The threshold ensures that items are seen at most a limited number of times in the training set.}
    \label{fig:cold_start}
\end{figure}

\subsection{Semantic ID Conflict Proportion Study}
We study how semantic ID conflict proportions, controlled through adjustments to codebook size and code depth, affect retrieval performance. The results are shown in Figure~\ref{fig:conflict_prop}. The codebook sizes for the different proportions are 512, 256, and 192, respectively. With a low proportion, our methods maintain stable gains, demonstrating the effectiveness of ECM and RRS. However, as the proportion becomes too high, the improvements begin to decline, suggesting that excessive latent conflicts introduce ambiguity in token mappings and impair quality. To achieve better performance, it is important to maintain semantics in the base codebooks, which allows our methods to fully realize their potential.

\subsection{Cold-Start Item Analysis}

Empirical evidence shows that semantic ID–based generative models perform poorly when recommending cold-start items~\cite{ding2024specgr,yang2024unifying}. Existing efforts to mitigate this issue typically rely on heuristic strategies~\cite{rajput2023recommender} or incorporate dense retrieval models~\cite{ding2024specgr,yang2024unifying}. One main reason is that existing semantic IDs often incorporate an additional non-semantic token to prevent conflicts, but the non-semantic tokens of unseen items are unpredictable. In contrast, our proposed purely semantic indexing uses only semantically meaningful tokens for each item. This allows us to investigate whether our indexing approach can improve cold-start performance without altering the model architecture.

We evaluate models on a subset of the test set containing items that appear at most $0$ or $1$ time.
The average performance over $3$ trials is illustrated in Figure~\ref{fig:cold_start}. We observe that both ECM and RRS improve performance on cold-start items, showing that our methods are effective for items with very limited interaction history. The improvements are especially notable for items that never appeared during training, demonstrating that purely semantic IDs enhance generalization to real-world cold-start scenarios.

\section{Conclusion}
In this paper, we address limitations in using non-semantic tokens to avoid semantic ID conflicts. We propose purely semantic indexing, prioritizing ID uniqueness over exact semantic feature reconstruction. We design two variants, exhaustive candidate matching (ECM) and recursive residual search (RRS), to generate unique, semantic-preserving IDs for ID construction methods with centroid-based codebooks.  Empirical results across tasks demonstrate our performance improvements and effectiveness in cold-start scenarios.

\bibliography{ref}
\end{document}